\documentclass[12pt]{article}
\usepackage{osajnl}
\begin{document}

\title{Saturated-absorption spectroscopy: Eliminating crossover
resonances using co-propagating beams}
\author{Ayan Banerjee}
\author{Vasant Natarajan}
\email{vasant@physics.iisc.ernet.in}
\affiliation{Department of Physics, Indian Institute of 
Science, 
Bangalore 560 012, INDIA}

\begin{abstract}
We demonstrate a new technique for saturated-absorption 
spectroscopy using co-propagating beams that does not have 
the problem of crossover resonances. The pump beam is locked 
to a transition and 
its absorption signal is monitored while the probe beam is 
scanned. As the probe comes into resonance with another 
transition, the pump absorption is reduced and the signal 
shows a Doppler-free dip. We use this technique to measure 
hyperfine intervals in the $D_2$ line of $^{85}$Rb with a 
precision of 70 kHz, and to resolve hyperfine levels in the 
$D_2$ line of $^{39}$K that are less than 10 MHz apart.
\end{abstract}
\maketitle

Laser spectroscopy in atomic vapor is often limited by 
Doppler broadening. The standard method to overcome the 
first-order Doppler effect is to use saturated-absorption 
spectroscopy \cite{DEM82}. In this technique, the laser is 
split into a weak probe beam and a strong pump beam, and 
the two beams are sent in opposite directions through the 
vapor. Due to the opposite Doppler shifts, only the 
zero-velocity atoms (i.e.\ atoms moving perpendicular to the 
direction of the beams) interact with both beams. For these 
atoms, the stronger pump beam saturates the transition and 
the probe-absorption spectrum shows a Doppler-free ``dip'' 
at line center. Ideally, the linewidth of the dip is 
limited only by the natural linewidth of the transition.

The above analysis is correct for two-level atoms, i.e.\ 
where the other atomic levels are far away compared to the 
Doppler width. However, in most real atoms, there are 
several closely-spaced hyperfine levels within the Doppler 
profile. The presence of multiple levels is a problem in 
saturated-absorption spectroscopy and results in what are 
called ``crossover resonances''. These spurious resonances 
occur when the laser is tuned exactly midway between two 
transitions, so that for some velocity group, the pump 
drives one transition while the probe drives the other. 
Since both transitions start from the same ground level, 
the pump is still effective in reducing absorption from the 
probe. Crossover resonances are particularly problematic 
for high-resolution spectroscopy on two levels whose 
spacing is only slightly larger than the natural linewidth. 
In this case, the crossover resonance often swamps the real 
peaks and makes it impossible to resolve the two 
transitions. 

In this article, we demonstrate an alternate technique for 
saturated-absorption spectroscopy that does not have the 
problem of crossover resonances. The technique uses to 
advantage the multilevel structure of the atom in creating 
differential absorption between the pump and probe beams. 
The basic idea is to have the two beams {\it co-propagate} 
through the vapor, with the pump frequency fixed on one 
transition while the probe is scanned over the other 
transitions. Unlike in conventional saturated-absorption 
spectroscopy, both beams have roughly equal intensities, 
and it is the absorption of the {\it pump beam} that is 
monitored. Since the pump is locked to a transition, it is 
absorbed only by the zero-velocity atoms. The absorption 
signal remains constant as the probe is scanned since the 
probe is generally absorbed by a different velocity group. 
However, as the probe comes into resonance with another 
transition for the zero-velocity group, the absorption of 
the pump is reduced and the signal shows a Doppler-free 
dip. 

There are two primary advantages to our scheme. The first, 
as mentioned earlier, is the absence of crossover 
resonances. The second is that the signal appears on a flat 
background. This is different from conventional 
saturated-absorption spectroscopy where the probe absorption has an 
underlying Doppler profile which can affect the 
determination of the peak centers. We have applied this 
technique to spectroscopy on the $D_2$ lines of $^{85}$Rb 
and $^{39}$K. In the case of Rb, we measure the hyperfine 
interval between the $F=1$ and 2 levels in the $5P_{3/2}$ 
state with a precision of 70 kHz. In K, we are able to 
resolve all three hyperfine transitions in the $D_2$ line, 
which are completely merged in ordinary 
saturated-absorption spectroscopy.

The first set of experiments were done with a vapor of 
$^{85}$Rb atoms. The experimental schematic is shown in 
Fig.\ 1. The pump and probe beams are derived from two 
frequency-stabilized diode lasers \cite{BRW01} operating on 
the Rb $D_2$ line at 780 nm ($5S_{1/2} \leftrightarrow 
5P_{3/2}$ transition). The intensity in each beam is about 
10 $\mu$W/cm$^2$. The two beams co-propagate through a 
room-temperature vapor cell such that the angle between 
them is less than 10 mrad. The pump laser is locked to a 
hyperfine transition using saturated-absorption 
spectroscopy in another Rb vapor cell. The probe laser is 
scanned by double-passing through an acousto-optic 
modulator (AOM) and scanning the AOM frequency. The double 
passing is necessary to maintain directional stability of 
the beam, which can otherwise lead to systematic shifts of 
the peak positions. The probe intensity is also stabilized 
by adjusting the rf power into the AOM in a servo loop. 

A normal saturated-absorption spectrum for the $D_2$ line 
in $^{85}$Rb is shown in Fig.\ 2. The underlying Doppler 
profile has been subtracted using a second probe beam that 
does not interact with the pump. The spectrum is for 
transitions starting from the $F=2$ ground level. The 
$F'=1$ and 2 levels in the excited state are barely visible 
because they are only 30 MHz apart \cite{AIV77} and overlap 
with the crossover resonance in between. The linewidth of 
the peaks is 12--15 MHz compared to the natural linewidth 
of 6 MHz. The primary causes for the increased linewidth 
are power broadening due to the pump beam and a small angle 
between the counter-propagating pump and probe beams. In 
addition, the lineshape of the peaks depends crucially on 
the intensities in the two beams. At high pump intensities, 
optical-pumping effects and the effect of velocity 
redistribution of the atoms from radiation pressure leads 
to inversion of the peaks or distortion of their Lorentzian 
lineshape \cite{GRM89}. This is seen in the lower trace of 
Fig.\ 2 where the pump intensity has been increased by a 
factor of 2. The $F'=1$ and 2 peaks get distorted and the 
$F'=(1,2)$ crossover resonance becomes inverted. 

The dramatic improvement in the spectrum with our new 
technique is seen in Fig.\ 3. The pump laser is locked to 
the $F=2 \leftrightarrow F'=3$ transition while the probe 
laser is scanned over a frequency range of 50 MHz covering 
the $F'=1$ and 2 hyperfine levels. The pump-transmission 
signal shows two well-resolved peaks corresponding to these 
levels. The spectrum appears on a flat background and there 
is no crossover resonance in between. The peaks also have 
symmetric Lorentzian lineshapes with no significant pulling 
due to the neighboring peak.

There are two advantages to scanning the probe laser using 
an AOM instead of the grating angle that controls the 
optical feedback. The first is that it guarantees linearity 
of the scan since the voltage-controlled oscillator (VCO) 
that determines the AOM frequency has a linear transfer 
function. Secondly, the scan axis has absolute frequency 
calibration once the voltage-to-frequency transfer function 
of the VCO is known. This transfer function is readily 
measured using a frequency counter. We have used the 
advantage of a calibrated frequency scan to precisely 
determine the hyperfine interval in Fig.\ 3. By fitting 
Lorentzians to the two peaks, we determine the peak centers 
with a precision of 50 kHz. This yields the value of the 
interval to be 29.35(7) MHz. The result is in good 
agreement with earlier values of 29.26(3) MHz \cite{RKN03} 
and 29.30(3) MHz \cite{BDN03} obtained by us using other 
techniques. The high level of precision in the current work 
is a direct consequence of being able to resolve the two 
transitions without an intervening crossover resonance. One 
potential source of error in hyperfine measurements is a 
systematic shift in the lock point of the laser from peak 
center. However, in our scheme, this is not a problem 
because both pump and probe beams address the same velocity 
group. Thus a shift in the pump-laser lock point causes the 
entire set of hyperfine peaks to shift but the intervals do 
not change.

We have performed a second set of experiments with 
$^{39}$K. For this, the two diode lasers are chosen to 
operate on the $D_2$ line of K at 767 nm ($4S_{1/2} 
\leftrightarrow 4P_{3/2}$ transition). The probe laser does 
not pass through an AOM but is scanned by scanning the 
grating. The consequent variation in the direction is 
negligible for the results presented here. The beams 
co-propagate through an ultra-high vacuum (UHV) glass cell 
maintained at a pressure below $10^{-8}$ torr using an ion 
pump. K vapor is produced by heating a getter source 
\cite{SAES} with a current of 2.6 A. The UHV environment is 
necessary to minimize linewidth broadening due to 
background collisions. The getter source also gives us 
control over the amount of K vapor in the cell, which is 
optimized for obtaining the narrowest linewidth.

In Fig.\ 4, we show the spectrum for $F=2 \rightarrow F'$ 
transitions in $^{39}$K. With conventional 
saturated-absorption spectroscopy, we obtain a single peak containing 
all the transitions. The underlying Doppler profile is also 
seen. The lineshape of the peak is a convolution of 6 peaks 
lying within 30 MHz of each other \cite{AIV77}. Indeed, it 
is hard to identify the location of the peak center with 
respect to the actual hyperfine transitions. By contrast, 
the new technique shows the locations of the three levels 
quite clearly. While the linewidth is still large enough 
that the peaks overlap, there is no ambiguity in 
identifying their locations. 

In summary, we have demonstrated a technique for 
saturated-absorption spectroscopy in multilevel atoms using 
{\it co-propagating} pump and probe beams. The pump laser is locked 
to a transition so that it addresses only the zero-velocity 
atoms. The transmission of the pump shows Doppler-free 
peaks when the probe comes into resonance with another 
transition for the same zero-velocity atoms. The main 
advantages of the technique are the absence of crossover 
resonances and the appearance of the signal on a flat 
background. We demonstrate the power of this technique for 
hyperfine spectroscopy of closely-spaced levels by using an 
AOM to measure the smallest interval in the $5P_{3/2}$ 
state of $^{85}$Rb. We are also able to resolve hyperfine 
transitions in the $D_2$ line of $^{39}$K. In the current 
work, we have used two lasers to generate the pump and 
probe beams, which appears disadvantageous compared to the 
use of one laser in conventional saturated-absorption 
spectroscopy. However, it is possible to generate the probe 
beam from the pump laser using one or more AOMs, which has 
advantages for scanning the probe in a calibrated manner. 
Finally, the pump intensity in our scheme has to be kept 
very small. At higher intensities, we see clear evidence of 
dressed states created due to the coherent driving by the 
pump laser. Indeed, we have used this technique as a means 
of studying dressed states in room-temperature vapor 
without complications from Doppler broadening \cite{RAN03}. 

The authors thank Dipankar Das for help with the 
experiments. This work was supported by the Department of 
Science and Technology, Government of India.


\newpage
\begin{figure}
\scalebox{0.47}{\includegraphics{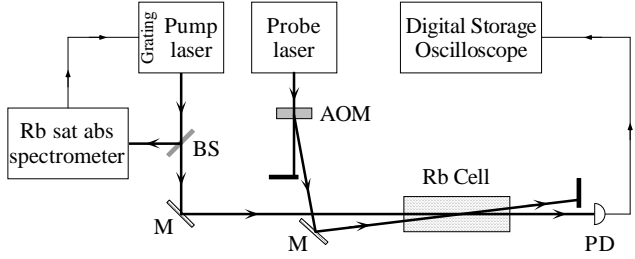}}
\caption{
Schematic of the experiment. The pump laser is locked to a 
hyperfine transition in Rb, while the probe laser is 
scanned by double passing through an AOM and scanning the 
AOM frequency. The two beams co-propagate through a Rb 
vapor cell. The angle between them has been exaggerated for 
clarity.
}
\label{f1}
\end{figure}

\begin{figure}
\scalebox{0.58}{\includegraphics{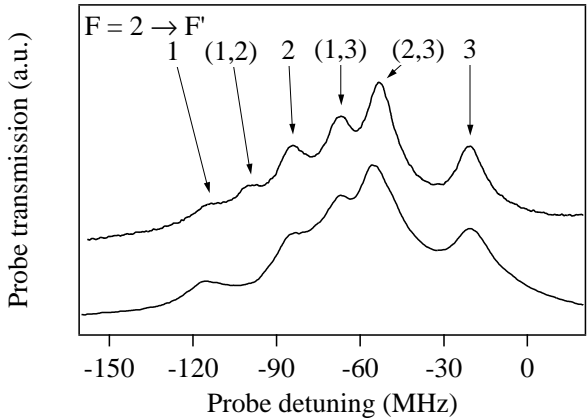}}
\caption{Saturated-absorption spectrum in the $D_2$ line of 
$^{85}$Rb for $F=2 \rightarrow F'$ transitions. The 
underlying Doppler profile has been subtracted. The 
transitions are labeled with the value of $F'$ and 
crossover resonances with the two values of $F'$ in 
brackets. The upper trace is the correct spectrum, while 
the lower trace is obtained when the pump intensity is 
increased by a factor of 2, resulting in inversion of the 
$F'=(1,2)$ peak. Probe detuning is measured from the 
unperturbed state.
}
\label{f2}
\end{figure}

\begin{figure}
\scalebox{0.58}{\includegraphics{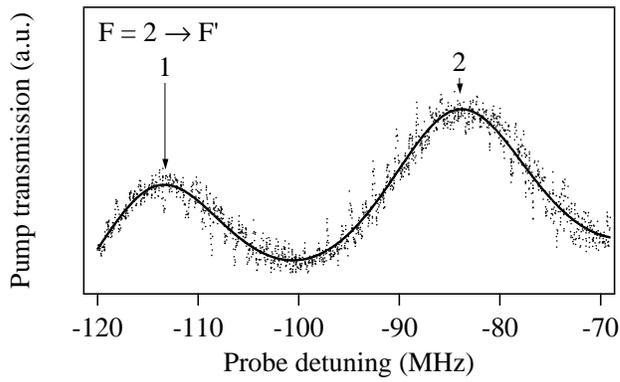}}
\caption{$^{85}$Rb spectrum obtained with new technique. 
The transitions are the same as in Fig.\ 2, with a narrow 
scan around the $F'=1$ and 2 levels. The frequency scale on 
the $x$-axis is set by the voltage-controlled oscillator 
driving the AOM. The Lorentzian fit (solid line) yields a 
value of 29.35(7) MHz for the hyperfine interval.
}
\label{f3}
\end{figure}

\begin{figure}
\scalebox{0.58}{\includegraphics{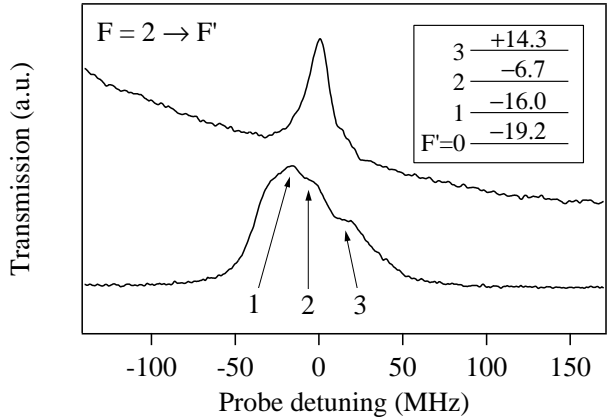}}
\caption{Saturated-absorption spectrum in the $D_2$ line of 
$^{39}$K. The upper trace shows the probe-transmission 
signal for $F=2 \rightarrow F'$ transitions in usual 
saturated-absorption spectroscopy. The 6 peaks are merged 
into a single peak since the hyperfine levels lie within 30 
MHz, as shown in the inset. The values in the inset are the 
frequency offset (in MHz) of each hyperfine level from the 
unperturbed state. The lower trace is the pump-transmission 
signal obtained with the new technique, clearly showing the 
locations of the hyperfine levels.
}
\label{f4}
\end{figure}

\end{document}